\documentclass[preprint,amssymb]{revtex4}
\usepackage{amsfonts}
\usepackage{amssymb}
\usepackage{slashed} 
\usepackage{graphicx}
\usepackage{dcolumn}

\usepackage{bm}
\def\be{\begin{equation}} 
\def\ee{\end{equation}}
\def\bea{\begin{eqnarray}} 
\def\eea{\end{eqnarray}}
\def\line{\hbox to \hsize}    
\def\frac #1#2{{#1\over #2}}

\def\tr{{\rm  tr\,}}

\def\psid{\psi^{\dagger}}

\def\Det{{\rm Det\,}}

\def\vev #1{{\langle #1\rangle}}
\def\1{\mbox{\bf 1}}
\def\bm#1{\mbox{\boldmath$#1$}} 

\begin{document}

\title{Effective action and electromagnetic response  of  topological superconductors and Majorana-mass Weyl fermions}

\author{MICHAEL STONE}

\affiliation{University of Illinois, Department of Physics\\ 1110 W. Green St.\\
Urbana, IL 61801 USA\\e-mail: m-stone5@illinois.edu}   

\author{PEDRO L.\ e S.\ LOPES}

\affiliation{Instituto de F\'isica Gleb Wataghin, Universidade Estadual de Campinas, Campinas, SP 13083-970, Brazil
\\e-mail: pedrolslopes@gmail.com}

\begin{abstract}  

Motivated by an apparent  paradox  in  [X-L.\ Qi, E.\ Witten, S-C.\ Zhang,
Phys.\ Rev.\ B {\bf 87} 134519  (2013)] 
we use the method of gauged Wess-Zumino-Witten functionals to construct an effective action for  a Weyl fermion whose Majorana mass   arises from coupling to a charged condensate.  We obtain expressions for the  current induced by an external gauge field and observe that the  topological part of the current is only one-third of that that might have been expected from the gauge anomaly. The anomaly is not changed  by the induced mass gap however.  The   topological  current is supplemented  by a  conventional supercurrent   that   supplies  the remaining two-thirds of the  anomaly once     the equation    of motion for the  Goldstone mode is satisfied.  We apply our formula for the current  to resolve the apparent  paradox, and also   to  the  chiral magnetic effect (CME) where it   predicts a  reduction of the CME current to   one third of its value for a free Weyl gas in thermal equilibrium. We attribute this reduction to a partial cancelation of the CME  
by a chiral vortical effect (CVE) current arising from the persistent rotation of the fluid  induced by the external magnetic field. 

 \end{abstract}

\maketitle

\section{Introduction}

In  \cite{qi-zhang-witten} Qi, Zhang and Witten (QZW) consider the electromagnetic response of  a 3+1 dimensional topological superconductor in which two   Fermi surfaces of  opposite Chern number are each provided with their  own  independent superconducting order parameter. 
When  the Fermi surfaces   are  realized as a pair of opposite-chirality Weyl fermions, the superconducting gap induced by the  order parameter is an example of   Majorana mass generation similar to that proposed for standard-model neutrinos.  The analysis in \cite{qi-zhang-witten}  therefore has potential applications well  beyond condensed matter physics.

One of  the intriguing  topological effects deduced   by QZW applies when one  (but not both) of the two condensate order parameters contains  a vortex line  about   which  the phase of the of charged condensate winds though $2\pi$.  If   an electric field is directed along the vortex line they find    an inflow of electric charge  into the vortex core. This  inflowing current  is similar to that which appears for vortex strings  in an uncharged Higgs field that induces a  mass for conventional Dirac fermions. In the Dirac case the  inflowing charge  is soaked up by the ${\rm U}(1)$  anomaly of a 1+1 dimensional charged chiral   fermion mode that is bound in the vortex core. Indeed the Dirac case  is the simplest illustration  of the Callan-Harvey anomaly-inflow mechanism \cite{callan-harvey}.       

For our Majorana-mass  Weyl fermion, the charge inflow  poses something of a paradox.   The vortex core still confines a 1+1 chiral fermion --- indeed  a Weyl fermion gapped by a charged Higgs field is a system  for which a  low  energy chiral vortex-core mode is guaranteed by the  Erick Weinberg   index theorem \cite{jackiw-rossi,weinberg-index}   --- however the chiral mode is a chiral-{\it Majorana\/} mode. It is  electrically  neutral (see appendix \ref{AP:vortex}) and hence   possess  no anomaly that can  absorb the inflowing current. 

This paradox leads us to reconsider the derivation of the effective action in \cite{qi-zhang-witten}. We  follow the route   pioneered in \cite{witten-global}  and seek an effective action that contains as its degrees of the freedom the ungapped phase degrees of freedom on the two Fermi surfaces. These Goldstone modes are then coupled to the external gauge field through the simplest set of interaction  terms that are compatible with the anomalous realization of the gauge symmetry.  The result is an action functional that is rather different from that obtained in  \cite{qi-zhang-witten} and enables us to resolve the inflow paradox.

In section \ref{SEC:Weyl} we describe  how  adding a  Majorana mass  to a charged Weyl fermion turns it into a superconductor, and  note   a second  potential  paradox that this threatens.
In section \ref{SEC:WZW} we review the strategy for systematically constructing the Wess-Zumino-Witten (WZW) effective action for the chiral dynamics of  anomalous system. We  apply this strategy to two-dimensional systems of charge density waves and superconductors  in section \ref{SEC:2d-applications} and demonstrate that   it reproduces familiar physics.   The more complicated case of four dimensions is addressed in section \ref{SEC:four-dimensions} where we find the topological currents and equation of motion for a superconducting Weyl fermion.  Equations (\ref{EQ:main-current}) and (\ref{EQ:main-EofM}) of this section are the principal results of this paper. In section \ref{SEC:discussion} we  show how our expression for the current resolves the inflow paradox, and also discuss the implications of these equations  for  the chiral magnetic effect. Finally in \ref{SEC:conclusions} we summarize and contrast our results with those of \cite{qi-zhang-witten}.

\section{Weyl fermions, superconductivity and  Majorana mass} 
\label{SEC:Weyl}

The prototype   of a system whose  Fermi surface possesses  a   non-trivial Berry connection    is a 3+1 dimensional Weyl fermion, where the  first Chern numbers of the Berry curvature  are  $C_1=\pm 1$ for right- and left-handed particles respectively.    

The second-quantized Hamiltonian for a  right-handed  Weyl particle with charge $e$  and coupled to an external Maxwell field $A^\mu= (\phi, {\bf A})$,  $A_\mu= (\phi, -{\bf A}) $ is
\bea
\hat H_{\rm Weyl}[A]&=& \int \psid\left\{  {\bm \sigma}\cdot ({\bf p}-e{\bf A})+e\phi \right\}\psi\, d^3x \nonumber\\
&=& \int \psid\left\{ -i {\bm \sigma}\cdot (\nabla-ie{\bf A})+e\phi\right\}\psi\, d^3x. 
\eea
The  anti-commuting two-component Fermi fields  $\psi$, $\psi^*$  obey  the canonical anti-commutation rules (CAR)
\be
\{\psi_\alpha(x),\psi^*_\beta(x')\}= \delta_{\alpha\beta}\,\delta^3(x-x'),
\ee
where  by  $\psi^*$, we mean the Hermitian conjugate of $\psi$ as a  Hilbert-space operator, but not a matrix transpose of the   two-component column spinor into a two-component row spinor. Thus
\be
\psi= \left(\matrix{ \psi_1\cr\ \psi_2}\right), \quad \psid= \left(\matrix{ \psi^*_1& \psi^*_2}\right).
\ee

The  field $\psi_{\rm c} =i{\sigma}_2\psi^*$   possesses identical  Lorentz
 transformation properties as $\psi$,  and also obeys the same CAR:
\be
\{\psi_{c,\alpha}(x),\psi^*_{c,\beta}(x')\}= \delta_{\alpha\beta}\delta^3(x-x').
\ee
The Pauli-matrix identity
\be
(i\sigma_2)^\dagger{\bm \sigma} (i\sigma_2) = -{\bm \sigma}^*
\ee
and an  integration  by parts permits us to  rewrite $\hat H_{\rm Weyl}$ in terms of $\psi_c$ as   
\bea
\hat H_{\rm Weyl} [A]= \int \psid_{\rm c} \left\{-{\bm \sigma}\cdot ({\bf p}+e{\bf A})-e\phi\right\}\psi_{\rm c}\, d^3x. 
\eea
The rewrite shows that $\psi_c$ is the charge conjugate (antiparticle) field of  $\psi$. The  original field  $\psi$   annihilates  positive energy   states that have charge $+e$ and are right handed in that the spin $ {\bm \sigma}$ is parallel to  ${\bf p}$.   The field  $\psi_{\rm c}$ annihilates   positive energy particles of   charge $-e$ that are {left}-handed in that their spin is \underline{anti}-parallel  to ${\bf p}$. 

The skew symmetry of $(i\sigma_2)_{\alpha\beta}= \epsilon_{\alpha\beta}$ allows  us to add  to $\hat H_{\rm Weyl}$ a  term 
\bea
\hat H_1 =\frac 12 (\Phi \psid \psi_{\rm c} + \Phi^* \psid_{\rm c} \psi )= \frac 12(\Phi \epsilon_{\alpha \beta}\psi^*_\alpha \psi^*_\beta -\Phi^*\epsilon_{\alpha\beta}\psi_\alpha\psi_\beta)
\label{EQ:Weyl-gap}
\eea
that   couples pairs of particles or antiparticles to  a charged  $c$-number Higgs field $\Phi=|\Phi| e^{i\theta} $.  We can write the resultant Hamiltonian in  Bogoliubov-de-Gennes (BdG) form  \cite{BdG}  as  
 \be
\hat H_{\rm BdG}[A] =\frac 12  \int d^3x \left(\matrix{\psid &\psid_{\rm c}}\right) \left[\matrix{ {\bm \sigma}\cdot ({\bf p}-e{\bf A}) +e\phi& \Phi\cr
                                           \Phi^* & -{\bm \sigma}\cdot ({\bf p}+e{\bf A})-e\phi}\right] \left(\matrix{\psi \cr \psi_{\rm c}}\right).
\label{EQ:BdG}
\ee
 The factor of 1/2 outside the integral ensures that  $\hat H_{\rm BdG} \to \hat H_{\rm Weyl}$ as  $\Phi\to  0$.

Were  our Weyl field   electrically neutral,  $\hat H_{\rm BdG }$ would 
contain the  one-particle  four-component Dirac hamiltonian 
\be
H_{\rm Dirac} = \left[\matrix{ {\bm \sigma}\cdot {\bf p}& \Phi\cr
                                           \Phi^* & -{\bm \sigma}\cdot {\bf p}}\right], 
\ee 
and  the added term would be  a  {\it Majorana mass\/}.  A   Majorana mass term  opens a gap at the ${\bf p}=0$ Dirac  point, but couples the right-handed particle to its own left-handed antiparticle rather than to an independent left-handed Weyl fermion.  It is as yet uncertain  whether   the  masses  of   standard-model  neutrinos arise from Dirac or Majorana terms \cite{avignone}.
The presence  of the gauge field $A^\mu$, however,   reveals a key  difference between the matrix appearing in (\ref{EQ:BdG})  and  the conventional one-particle   Hamiltonian for charged Dirac particles
\be
 H_{\rm Dirac }(A) =\left[\matrix{ {\bm \sigma}\cdot ({\bf p}-e{\bf A}) +e\phi& \Phi\cr
                                           \Phi^* & -{\bm \sigma}\cdot ({\bf p}-e{\bf A})+e\phi}\right].
\ee

The sign-difference in the $(\phi, {\bf A})$  coupling to the left- and right-handed fields  in (\ref{EQ:BdG}) means  that the   BdG  fermions  are coupled to an  {\it axial-vector\/} gauge field. The four-dimensional Lagrange density is therefore of the form 
\be
{\mathcal L}_{\rm BdG}=  \bar\Psi  \left(i \gamma^\mu (\partial_\mu +ie\gamma_5 A_\mu)-|\Phi| e^{i\gamma_5 \theta} \right)\Psi.
\label{EQ:axialDirac}
\ee
One consequence of the axial character of the gauge field  appears 
when  we set ${\bf A}=0$ and let   $-e\phi=-eA_0$ be regarded as   a chemical potential $\mu$ that fixes  the Fermi-surface of the  Weyl fermions  to be at $|{\bf p}|=\mu$.  We find that the  energy gap in the spectrum of $\hat H_{\rm BdG}[A] $  appears  at this  {Fermi surface\/} rather than at the Dirac point ${\bf p}=0$.   Consequently  a degenerate gas   of  gauge-coupled  neutrinos with a Majorana mass is really  a topological superconductor \cite{wilczek}.

A second and potentially  paradoxical issue is the question of the gauge  anomaly. The original massless Weyl fermion possesses an anomaly in the conservation law  for  the particle number  current $j^\mu= (\psi^\dagger\psi, \psi^\dagger {\bm \sigma} \psi)$  that modifies it to read 
\bea
\partial_\mu j^\mu= \frac{e^2}{32\pi^2} \epsilon^{\mu\nu\sigma\tau}F_{\mu\nu}F_{\sigma\tau}.
\label{EQ:right-weyl-anomaly}
\eea
The  non-conservation of charge arises from a flux of particles from the negative-energy  Dirac sea (regarded as a charge-neutral vacuum)  into the positive-energy Fermi sea {\it via\/} the Dirac point at a rate 
$\dot N= e^2 {\bf E}\cdot {\bf B}/4\pi^2$ per unit volume.  This anomaly is no trouble for   the gauge invariance of our   topological superconductor as each Fermi surface with a positive Chern number is paired with one with a negative Chern number and hence a canceling  anomaly. The entire theory  is therefore  {anomaly-free\/}. What {\it is\/}  a potential problem is that it is  that the axial-current anomaly for  Dirac particle coupled to a {\it axial-vector\/}  gauge field is only 1/3 that of the axial current anomaly for  
Dirac particle coupled to a conventional  vector gauge field \cite{gross-jackiw, bertlmann}.    The 1/3  is a puzzle because  a  mere cosmetic rewrite combined with  the  introduction of a mass term (a ``soft'' low-energy peturbation) should not be able to alter an  anomaly  that    arises from high-energy effects.  Note also that  compared to the usual axial anomaly,  there is an additional factor of $1/2$ to be taken into account in  the current because the fermi field in  (\ref{EQ:axialDirac}) obeys a Majorana condition $\Psi=C\Psi^*$, but this is already included in (\ref{EQ:right-weyl-anomaly}) because we have only right-handed particles. The usual Dirac-particle axial anomaly counts the difference between the number of right handed and left handed particles and so has a RHS  that is  twice that of (\ref{EQ:right-weyl-anomaly}). 

In the next to section we will set out to resolve the two potential paradoxes by investigating  the form of the effective action 
\be
S[\theta, A] =-\frac i2 \ln \Det  \left(i \gamma^\mu (\partial_\mu +ie\gamma_5 A_\mu)-|\Phi| e^{i\gamma_5 \theta} \right)
\label{EQ:determinant}
\ee
that arises from integrating out the right-handed  Weyl fermion.  In (\ref{EQ:determinant}) the factor of 1/2 comes from the Majorana/BdG  condition.

\section{Wess-Zumino effective actions}
\label{SEC:WZW}

We begin by reviewing   Witten's Wess-Zumino strategy  \cite{witten-global}  that  enables us to deduce, with minimal labour,  the topological part of the  effective action for a  massive  fermion  coupled to a gauge field.  To appreciate  the underlying structure of the method  we  first consider  the general case of  $N$ flavours of   fermions $\psi_L$, $\psi_R$ that  are coupled to  non-Abelian ${\rm U}(N)_L\times {\rm U}(N)_R$  gauge fields. Only then  will we restrict ourselves to   fermions obeying  $\psi_L=(\psi_R)_c$ and  Abelian  axial-vector gauge fields. 

The action for the  fields $\psi_R$, $\psi_L$ is   built from the gauge covariant derivatives 
\bea
\nabla_\mu \psi_R &=& (\partial_\mu+R_\mu)\psi_R,\nonumber\\
\nabla_\mu \psi_L &=& (\partial_\mu+L_\mu)\psi_L.
\eea
We have here absorbed the  customary    factors of $i$  and $e$ into the  definition of the gauge potentials $R$ and $L$ so as to improve the readability of our formul\ae. These factors will  be  restored when we consider physical effects.
We  also sometimes write $\Psi=(\psi_R,\psi_L)^T$ and  
\be
\nabla \Psi= (\partial +\mathcal {V}+\gamma_5 {\mathcal A})\Psi,
\ee
where the vector and axial-vector  gauge fields ${\mathcal V}$ and $\mathcal A$ are related to the right and left gauge fields by  $ \mathcal {V}+{\mathcal A}=R$ and $ \mathcal {V}- {\mathcal A}=L$.

Our fermions are  gapped by coupling to a nonlinear $\sigma$-model field $U\in {\rm U}(N)$ 
through  a term
 \be
H_{\rm mass}=\Delta( \psi^*_{L,i} U_{ij} \psi_{R,j} +   \psi^*_{R,i} U^\dagger_{ij} \psi_{L,j}).
\ee
 This  form for the gap-inducing interaction  makes  the $\Phi$ appearing in  (\ref{EQ:Weyl-gap}) and (\ref{EQ:BdG})  correspond  with $U^\dagger$ rather than $U$, but we have  adopted it so as to facilitate comparison  with the notation in \cite{witten-global}. 

The resulting classical action is invariant under 
 the transformation  $(h_L,h_R)\in {\rm U}(N)_L\times {\rm U}(N)_R$ that acts to take   $\psi_R\to h^{-1}_R \psi_R$, $\psi_L \to h^{-1}_L \psi_L$, $U\to h_L^{-1} U h_R$,
while the  gauge-potential  1-forms $L$, $R$ and their associated  curvature 2-forms $F_L=dL+L^2$ and $F_R=dR+R^2$ transform as 
\bea
L&\to& L^{h_L}=h_L^{-1} Lh_L +h_L^{-1}d h_L,\nonumber\\
R&\to& R^{h_R}= h_R^{-1} Rh_R +h_R^{-1}d h_R,\nonumber\\
 F_L&\to& h_L^{-1} F_L h_L,\nonumber\\
  F_R &\to& h_R^{-1} F_R h_R.
 \label{EQ:classical-gauge} 
\eea
 This gauge invariance will be partially violated in the quantum theory by anomalies.  
 
 The transformation rules (\ref{EQ:classical-gauge}) show that the appropriate   covariant derivative for the non-linear $\sigma$-model field  $U$ is 
 \be
 \nabla_\mu U = \partial_\mu U +L_\mu U - UR_\mu.
 \ee 
 This derivative transforms in the same manner as $U$ itself 
 \be
 \nabla _\mu U \to h_L^{-1} (\nabla_\mu U) h_R.
 \ee

Because the fermions are fully gapped, they are slaved to the external gauge and mass-generating fields. Their response to adiabatic changes in those fields  is therefore governed by an effective action $S[R,L,U]$ which contains a topological part,  the gauged Wess-Zumino-Witten (WZW) action $W[R,L,U]$,
that is entirely determined by   anomalies \cite{wess-zumino,witten-global}.   The functional    $W[R,L,U]$ can be systematically constructed by imagining that our four-dimensional theory lives on the boundary of a five manifold (the ``bulk'')  from which a current inflow is the source of the anomalous conservation laws  \cite{callan-harvey}.    The action in the bulk will involve a  five-dimensional Chern-Simons form constructed so that the  complete  topological  action functional (bulk-plus-boundary)  is fully gauge invariant.

When the bulk is $2n-1$ dimensional, its   Chern-Simons action density $\omega_{2n-1}(L,R)$   is to be a solution to
\be
d \omega_{2n-1}(R,L)= \Omega_{2n}(F_R)-\Omega_{2n}(F_L),
\label{EQ:CS}
\ee
where $\Omega_{2n}(F)= \tr\{F^n\}$ is the (unnormalized) Chern-character anomaly polynomial. The minus sign between the two terms reflects the 
fact that left and right handed fermions have opposite anomalies.
An  obvious  way to satisfy  (\ref{EQ:CS}) would be to set  
\be
\omega_{2n-1}(R,L) = \omega_{2n-1}(R)-\omega_{2n-1}(L),
\ee
where $\omega_{2n-1}(R)$, $\omega_{2n-1}(L)$ are the standard  Chern-Simons $(2n-1)$-forms for a single gauge field. For example
\bea
\omega_3(A)&=&\tr\left\{AF-\frac 13 A^3,\right\},\nonumber\\
&=& \tr\left\{AdA+\frac 23 A^3\right\};\\
\omega_5(A)
&=& \tr\{ A(dA)^2 +\frac 32 A^3 dA +\frac 35 A^5\},\nonumber\\
&=& \tr\{ AF^2-\frac 12 FA^3 +\frac 1{10} A^5\}.
\eea

For constructing a  Wess-Zumino-Witten action with a {\it single\/}  non-linear $\sigma$-model field $U$  --- as opposed to one with separate fields $g_L\in {\rm U}(N)_L $ and  $g_R \in {\rm U}(N)_R$ ---  it is necessary  to have a solution $\tilde\omega_{2n-1}(L,R)$ to (\ref{EQ:CS})  that is invariant under a diagonal $(h_R=h_L)$ gauge transformation, {\it i.e.\/}
\be
\tilde \omega_{2n-1}(R^h,L^h)=\tilde \omega_{2n-1}(R,L).
\ee
How to arrange for  this is shown by Ma{\~n}es \cite{manes}. His  idea is to consider
\be
\Omega_{2n}(t)\stackrel{\rm def}{=}\Omega_{2n}(F_{+,t})-\Omega_{2n}(F_{-,t})
\ee
where
\be
A_{+,t}= tR+(1-t)L,\quad A_{-,t}= tL+(1-t)R.
\ee
Then $\Omega_{2n}(1)= \Omega_{2n}(F_R)-\Omega_{2n}(F_L)=-\Omega_{2n}(0)$, so
\be
2\Omega_{2n}(1)=\Omega_{2n}(1)- \Omega_{2n}(0)=\int_0^1 \partial_t \Omega_{2n}(t)\,dt.
\ee
The  transgression formula  for the variation of the Chern character gives
\be
\partial_t  \Omega_{2n}(t)=n\,d\left( \tr\{(R-L)F_{+,t}^{n-1}\}- \tr\{ (L-R)F_{-,t}^{n-1}\}\right)
\ee 
and  we can  take   
\be
\tilde \omega_{2m-1}(R,L)=\frac n2 \int_0^1 \left( \tr\{(R-L)F_{+,t}^{n-1}\}- \tr\{ (L-R)F_{-,t}^{n-1}\}\right) dt.
\label{EQ:manes-solution}
\ee
Under a diagonal gauge transformation the inhomogeneous terms cancel so that  $(R-L)\to h^{-1}(R-L)h$, and $F_{\pm,t}\to h^{-1}F_{\pm,t}h$. Consequently the integrand in  (\ref{EQ:manes-solution})  is manifestly  invariant under this transformation  and the invariance is inherited by   $\tilde\omega_{2m-1}(R,L)$. 
Solutions to (\ref{EQ:CS}) can differ only by  the $d$ of something, and so 
\be
\tilde \omega_{2m-1}(R,L) =   \omega_{2n-1}(R)-\omega_{2n-1}(L)+d S_{2n-2}(R,L)
\ee
for some $S_{2n-2}(L,R)$. 

We now define
\be
\tilde C[R,L]= \frac {i^n}{(2\pi)^{n-1}n!}\int_{M_{2n-1}} \tilde \omega_{2n-1}[R,L],
\ee
where the normalization has been  chosen so as to reproduce the perturbation-theory anomaly.  We have  
\be
\tilde C[R^{g_R},L^{g_L}] = \tilde C[R, L^{g_Lg^{-1}_R}]=\tilde C[ R,L^U],
\ee
where $U=g_Lg^{-1}_R$. This allows us to   
 define the gauged  Wess-Zumino-Witten  functional $W[R,L,U]$ by  setting 
 \be
 \tilde C[ R,L^U]= C[R,L] +W[R,L,U],
 \ee
 where 
  \be 
  C[R,L]= \frac {i^n}{(2\pi)^{n-1}n!}\int_{M_{2n-1}} \omega_{2n-1}[R,L],
  \ee
  as $S_{2n-2}[R,L^U]$  has no $U$-independent part.  Variations  of the functional $W[R,L,U]$ depend only on the values of the  fields $L$, $R$ and $U$ on the boundary of $M_{2n-1}$, and so $W$ can serve as an action on the  $2n-2$ dimensional space-time $M_{2n-1}=\partial M_{2n-1}$. 
 The functional  $\tilde C[R^{g_R},L^{g_L}] $  has been constructed to be  invariant under 
 \bea
 R&\to& R^{h_R}=h_R^{-1}Rh_R  +h_R^{-1}dh_R, \nonumber\\
 L&\to& L^{h_L}=h_L^{-1}Lh_L +h_L^{-1}dh_L,\nonumber\\
 U&\to& h_L^{-1}Uh_R.\
 \eea
 which coincides with (\ref{EQ:classical-gauge}). The equivalent functional   $\tilde C[ R,L^U]$ is therefore also gauge invariant. Its bulk Chern-Simons  and boundary Wess-Zumino functionals, $C[R,L]$  and $W[R,L,U]$ respectively,  are not separately gauge invariant. The gauge dependence of $W[R,L,U]$ is the source of the anomaly. 
 
A  key ingredient in $W[R,L,U]$ is  the $(2n-2)$-form     $S_{2n-2}$.  This form    is  the  ``Bardeen counterterm''  that was originally introduced  by W.~Bardeen  to ensure that the ``consistent anomaly'' vanished for vector currents. Here it must be included in the action  for a rather different reason:   the left and right Dirac seas are being glued together by the single mass-generating field multiplet $U$ \cite{manes}.

 \section{Two-dimensions: application to  superconductors and charge-density waves} 
 \label{SEC:2d-applications}
 
 \subsection{Currents and anomalies}
 
As an illustration of the WZW strategy consider a theory on  a two space-time dimensional surface $M_2$ that is the boundary of a   three-dimensional bulk $M_3$. In this case  case $n=2$ and Ma\~nes' construction  gives 
\be
\tilde\omega_3(R,L)= \omega_3(R)-\omega_3(L) +d\,\tr\{LR\}.
\ee
We can verify the diagonal invariance by using  
\be
\omega_3(A^g)=\omega_3(A) -\frac 13 \tr\{(g^{-1}dg)^3) - d \tr\{dg g^{-1} A\}
\ee
to find that
\bea
\tilde\omega_3(R^g,L^g)-\tilde \omega_3(R,L) 
&=& -\tr\{dgg^{-1}R\} +\tr\{ dg g^{-1}L\}\nonumber\\
&&\quad +\tr \{Ldg g^{-1} \} + \tr\{dg g^{-1}R \}+ \tr\{(g^{-1}dg)^2\}\nonumber\\
{}&=&0
\label{EQ:verify}
\eea
We have taken note that  the last term in the penultimate line in (\ref{EQ:verify}) is zero.
The modified Chern-Simons action  is therefore  invariant under  {\it vector} gauge transformations even if $F_R\ne F_L$.  
 
The bulk-plus-boundary topological and gauge-invariant action is now  
 \bea
 \tilde C[R,L^U] &=&\frac{1}{4\pi} \int_{M_3} \tr\{\omega_3(R)-\omega_3(L^U)\} -\frac{1}{4\pi} \int_{M_2}\tr\{RL^U\} \nonumber\\
 &=& \frac{1}{4\pi} \int_{M_3} \tr\{\omega_3(R)-\omega_3(L)+\frac 13 (U^{-1}dU)^3\} \nonumber\\
 && +\frac{1}{4\pi} \int_{M_2} \tr\{dUU^{-1} L -RU^{-1} dU-RU^{-1}LU \}.\nonumber
 \eea
 We compute the  currents $J_R^\mu$ and $J_L^\mu$ that flow in $M_2$ from the boundary part of the variations of $\tilde C$  with respect to $R$ and $L$. These variations are  
 \bea
 \delta_R \tilde C&=& \frac 1{4\pi} \int_{M_2} \tr\{ \delta R(R-U^{-1}LU-U^{-1}dU\}\nonumber\\
 &=&    \frac 1{4\pi} \int_{M_2} \tr\{ \delta R(-U^{-1} \nabla U\}\nonumber\\
 &\equiv& i\int_{M_2}  d^2x \sqrt{g} \, \tr\{ \delta R_\mu J^\mu_R\},
 \eea
 and 
 \bea
 \delta_L \tilde C&=& \frac 1{4\pi} \int_{M_2} \tr\{ \delta L(-L-URU^{-1}-dUU^{-1}\}\nonumber\\
 &=&    \frac 1{4\pi} \int_{M_2} \tr\{ \delta L(- \nabla U U^{-1}\}.\nonumber\\
 &\equiv& i\int_{M_2}  d^2x \sqrt{g} \, \tr\{ \delta L_\mu J^\mu_L\} .
 \eea
  We  read off that
 \bea
 J^\mu_R&=& -\frac{\epsilon^{\mu\nu}}{4\pi i} \,U^{-1} \nabla_\nu U\nonumber\\
 J^\mu_L &=&-\frac{\epsilon^{\mu\nu}}{4\pi i}\, (\nabla_\nu U)U^{-1}.
 \label{EQ:2d-currents}
 \eea

 The  currents (\ref {EQ:2d-currents})  automatically include the Bardeen polynomial  terms  that convert  ``consistent''   currents  to ``covariant''  currents  \cite{bardeen-zumino-consistent-covariant}. The Bardeen polynomials   are here simply  the  integrated-out  boundary   parts of the variation of the  bulk  Chern-Simons action (see, for example, \cite{stone-righi}). In the absence of  the $M_3$  bulk  these polynomials    have to be motivated  and added by hand  as was done in  \cite{bardeen-zumino-consistent-covariant}.
 The currents being  \underline{covariant} means that    that under a gauge transformation $(h_R,h_L)$ each current transforms in the adjoint representation of its appropriate   group   
 \bea
 J_R^\mu &\to& h_R^{-1} J^\mu_R h_{R},\nonumber\\
 J_L^\mu &\to& h_L^{-1} J^\mu_L h_{L}.\nonumber
 \eea
 This property is easily verified.   As a consequence of the transformation properties of the currents their  appropriate covariant derivatives  are 
 \bea
 \nabla_\mu J^\nu_L= \partial_\mu J^\mu_L +[L_\mu, J^\nu_L],\nonumber\\
 \nabla_\mu J^\nu_R= \partial_\mu J^\mu_R +[R_\mu, J^\nu_R].\nonumber
 \eea
 
 While we are envisaging the gauge fields as being externally imposed, the non-linear $\sigma$-model field $U$ is autonomous.  In order for our   currents to satisfy their conservation laws  we need   $U$ to obey its   equation of motion. This we obtain by  setting to zero the variation of  $\tilde C[R,L^U]$ due to  an arbitrary  change in $U$. The variation  is the integral of 
\be
\frac{1}{4\pi}{\rm tr}\left\{U^{-1} \delta U\left((U^{-1}dU)^2 +[U^{-1}dU, U^{-1}LU]_+-U^{-1}dLU -dR -[R,U^{-1}dU+U^{-1}LU]_+\right)\right\},
\label{EQ:2d-EofM}
\ee
and gives the   matrix valued equation,
\bea
0&=&\frac{1}{4\pi}\epsilon^{\mu\nu}\left( U^{-1}\partial_\mu U U^{-1} \partial_\nu U +[U^{-1}\partial_\mu U,U^{-1}L_\nu U]-U^{-1}(\partial_\mu L_\nu) U\right. \nonumber\\
&& \left. \qquad -\partial_\mu R_\nu -[R_\mu, U^{-1}\partial_\nu U+U^{-1}L_\nu U]\right).
\label{EQ:2d-EofM2}
\eea
When  we substitute
\be
J^\mu_R = -\frac {\epsilon^{\mu\nu}}{4\pi i }U^{-1}(\partial_\nu U+L_\nu U-UR_\nu)
\ee
into 
\be
 \nabla_\mu J^\mu_R=\partial_\mu J^\mu_R +[R_\mu, J^\mu_R]
 \ee
and make  use of the equation of motion   we verify the covariant anomalous conservation law
 \be
 \nabla_\mu J^\mu_R= \frac{1}{4\pi i} \epsilon^{\mu\nu} F^R_{\mu\nu}.
 \label{EQ:2d-anomR}
 \ee
 A similar equation 
 \be
  \nabla_\mu J^\mu_L= -\frac{1}{4\pi i} \epsilon^{\mu\nu} F^L_{\mu\nu}
  \label{EQ:2d-anomL}
 \ee
holds for $J^{\mu}_L$.

When we restrict ourselves to an Abelian gauge group,  equation   (\ref{EQ:2d-EofM2}) reduces to $dL+dR=0$, which is a constraint on the external gauge fields rather than an equation of motion. This awkwardness  is resolved by remembering  that in addition to the topological terms  the complete action will contain non-topological but manifestly gauge invariant terms such as an  ordinary non-linear $\sigma$-model  action 
\bea
S_{\rm conventional}[U, R,L]&=& \frac {f^2}2 \int d^2 x \, g^{\mu\nu}\tr\{  \nabla_\mu U\nabla_\nu U^\dagger\}\nonumber\\
&=& -\frac {f^2}2 \int d^2 x  \,g^{\mu\nu}\tr\{ (U^{-1} \nabla_\mu U)(U^{-1}\nabla_\nu U)\}.
\label{EQ:nonab-nontop}
\eea
After including the  contribution from this action we recover  a proper equation of motion.

\subsection{Abelian applications}

In the Abelian case we can set  $U=e^{-i\theta}$  (recall that our $\Phi$ corresponds to   $U^\dagger$, so $\Phi=|\Phi|e^{i\theta}$) and restore the factors of $i$ so that $R= iR_\mu dx^\mu$, $L=iL_\mu dx^\mu$  then (\ref{EQ:nonab-nontop}) becomes 
 \be
 S_{\rm conventional}[\theta,R,L]=\frac {f^2}2 \int d^2 x  \{(\partial_\mu \theta -L_\mu+R_\mu) (\partial^\mu \theta -L^\mu+R^\mu)   \}.
\ee
The currents become 
\bea
 J^\mu _R&=& -f^2(\partial^\mu \theta -L^\mu+R^\mu)    +\frac{\epsilon^{\mu\nu}}{4\pi}(\partial_\nu \theta -L_\nu+R_\nu), \nonumber\\
 J^\mu_L &=&+f^2(\partial^\mu \theta -L^\mu+R^\mu)    +\frac{\epsilon^{\mu\nu}}{4\pi}(\partial_\nu \theta -L_\nu+R_\nu),
\eea
and the equation of motion for $\theta$  is modified to 
\be
- f^2 \partial_\mu (\partial^\mu \theta -L^\mu+R^\mu) - \frac{\epsilon^{\mu\nu}}{8\pi}(F^L_{\mu\nu}+ F^R_{\mu\nu})=0.
\label{EQ:2d-abelian-EofM}
\ee
The anomalous conservation laws  (\ref{EQ:2d-anomR}) and (\ref{EQ:2d-anomL}) remain unchanged.  

If we restrict ourselves to the case of vector gauge fields only ($L=R$, ${\mathcal A}=0$) the vector current ${J}_{V}^\mu \equiv  J^\mu _R+J^\mu _L$ reduces to  
\be
{J}_{V}^\mu  = \frac{\epsilon^{\mu\nu}}{2\pi}\partial_\nu \theta, 
\label{EQ:2d-vector-vector}
\ee
which  is  the   automatically-conserved  current found by  Goldstone and Wilczek \cite{goldstone-wilczek}.  The axial current 
\be 
{J}_{A}^\mu \equiv  J^\mu _R-J^\mu _L= - 2f_\pi^2\partial^\mu \theta
\ee
obeys 
\be
\partial_\mu{J}_{A}^\mu = \frac 1{2\pi} \epsilon^{\mu\nu} F^V_{\mu\nu},
\ee
(where $ F^V\equiv  F^L=F^R$) 
by virtue of the equation of motion (\ref{EQ:2d-abelian-EofM}) for $\theta$.   

This  vector  gauge-field case  provides a model for the   conductivity of  a sliding charge-density wave (CDW).  In a one-dimensional CDW  the Fermi  surface  is gapped by a potential 
\be
V(x,t)= \Delta \cos(2k_fx -\theta_{\rm CDW}(x,t))
\ee 
which  arises  from a Peierls distortion of the lattice and couples  the two Fermi points at $k=\pm k_f$. The electronic states near the Fermi energy are  described by a Hamiltonian
\be
\hat H_{\rm CDW}= \int dx \left(\matrix {\psi_R^* & \psi^*_L}\right) \left[\matrix{-iv_f (\partial_x-ieA_x) +e\phi & \Delta e^{-i\theta_{\rm CDW}}\cr \Delta e^{+i\theta_{\rm CDW}} & iv_f (\partial_x-ieA_x)+e\phi} \right] \left(\matrix{ \psi_R \cr \psi_L}\right)
\ee
that  is of Dirac form (and with a Dirac mass),  but with the speed of light replaced by the Fermi velocity $v_f$  so that $g_{\mu\nu}={\rm diag}(v_f^2,-1)$.
The resulting number density and current  are \cite{lee-rice-anderson}
\bea
J^0_V&=& \vev{\psi^*_R\psi_R+\psi^*_L\psi_L} = -\frac{1}{2\pi}\partial_x \theta_{\rm CDW}\nonumber\\
J^1_V&=& v_f  \vev{\psi^*_R\psi_R-\psi^*_L\psi_L} =+\frac{1}{2\pi}\partial_t \theta_{\rm CDW}
\eea
and coincide with  (\ref {EQ:2d-vector-vector}) once we notice that our  convention for the  CDW phase $\theta_{\rm CDW}$ gives it the  opposite sign to our previous $\theta$.  That the equation of motion for $\theta_{\rm CDW}$ provides a condensed-matter analogue of the axial anomaly was pointed out in  \cite{krive-rozhavsky,su-sakita}.

For us, a   more interesting case occurs when we specialize to axial gauge fields $A_\mu \equiv R_\mu=-L_\mu$.   Then the axial gauge current 
\be
J^\mu_A= -2f^2 (\partial^\mu \theta +2A^\mu))
\label{EQ:2d-axial-axial}
\ee
is conserved by the equation of motion 
\be
\partial_\mu J^\mu_A=\partial_\mu (-f^2(\partial^\mu \theta +2A^\mu))=0,
\label{EQ:2d-axial-conservation}
\ee
while the vector current 
\be
J_V^\mu =\frac{\epsilon^{\mu\nu}}{2\pi} (\partial_\nu\theta +2A_\nu)
\label{EQ:2d-axialfield-vectorcurrent}
\ee
is anomalous
\be
\partial_\mu J^\mu_V= \frac{\epsilon^{\mu\nu}}{2\pi} \partial_\mu (\partial_\nu\theta +2A_\nu)= \frac{\epsilon^{\mu\nu}}{2\pi} F^A_{\mu\nu}.
\label{EQ:2d-vector-anomaly}
\ee
Here $F^A\equiv F^R=-F^L$.

These equations are applicable to  a non-relativistic  1+1 dimensional BCS superconductor.   If we linearize near the Fermi surface,  the BdG    Hamiltonian becomes 
\be
\hat H_{\rm SC}= \int dx \left(\matrix {\psi_{\uparrow,R}^* & \psi_{\downarrow, L}}\right) \left[\matrix{-iv_f (\partial_x-ieA_x)+e\phi & \Delta e^{i\theta}\cr \Delta e^{-i\theta} & iv_f (\partial_x+ieA_x)-e\phi} \right] \left(\matrix{ \psi_{\uparrow, R} \cr \psi_{\downarrow, L}^*}\right),
\ee
 together with  another term for the opposite spin components.   
 
This Hamiltonian is again of Dirac form (again with a Dirac mass), but  involves  a two-component Nambu  spinor 
 \be
 \Psi=\left(\matrix{ \psi_{\uparrow, R} \cr \psi_{\downarrow, L}^*}\right)
 \ee 
 The physical number current is therefore  the axial current 
 \bea
 J^0_{\rm Num}&=&  : \Psi^\dagger \sigma_3  \Psi:\nonumber\\
 &=& :( \psi^*_{\uparrow, R} \psi_{\uparrow, R} -\psi_{\downarrow, L}\psi^*_{\downarrow, L}):\nonumber\\
 &=& \psi^*_{\uparrow, R} \psi_{\uparrow, R} +\psi^*_{\downarrow, L}\psi_{\downarrow, L}\\
 J^1_{\rm Num}&=& v_f : \Psi^\dagger  \Psi: \nonumber\\
 &=& v_f :( \psi^*_{\uparrow, R} \psi_{\uparrow, R} +\psi_{\downarrow, L}\psi^*_{\downarrow, L}):\nonumber \\
&=& v_f( \psi^*_{\uparrow, R} \psi_{\uparrow, R} -\psi^*_{\downarrow, L}\psi_{\downarrow, L}).
 \eea
 
In a superconductor the  ${\rm U}(1)$ particle-number symmetry  is 
spontaneously broken by the condensate. However   conservation of the  number current (here  the   axial current)  is recovered  once we impose the equation of motion for the condensate order parameter --- just as happens in (\ref{EQ:2d-axial-conservation}).

 What about the anomalous conservation law  (\ref{EQ:2d-vector-anomaly}) for the vector current?
 After multiplication by the Fermi momentum $k_f$ we can identify the normal-ordered   vector-current  density with the electron momentum density ${T^0}_1$, and its space component  with the momentum flux $ {T^1}_1$.  Thus 
 \bea
 {T^0}_1 &=&  k_f  (:\Psi^\dagger  \Psi:)=k_f (\psi^*_{\uparrow, R} \psi_{\uparrow, R} -\psi^*_{\downarrow, L}\psi_{\downarrow, L}),\nonumber\\
 {T^1}_1 &=& k_f  v_f (:\Psi^\dagger  \sigma_3 \Psi:) =k_f v_f (\psi^*_{\uparrow, R} \psi_{\uparrow, R} +\psi^*_{\downarrow, L}\psi_{\downarrow, L}).
 \eea
 By   using (\ref{EQ:2d-axialfield-vectorcurrent})  for the vector current we find 
 \bea
 {T^0}_1 &=& \frac{k_f}{2\pi} (\partial_1\theta +2eA_1)   \nonumber\\
 {T^1}_1 &=& - \frac{k_f}{2\pi} (\partial_0\theta +2eA_0),  
 \eea
 so  the anomaly in the vector current therefore reads 
 \bea
 \partial_0{T^0}_1 +\partial_1 {T^1}_1 &=&  \frac{k_f}{\pi} e (\partial_0A_1-\partial_1 A_0)\nonumber\\
 &=& \rho e E_1.
 \eea
 Here $\rho= k_f/\pi$ is the particle-number density, and we have used that in our $(+,-,\ldots)$ metric convention $A_1$ is {\it minus\/} the $x$ component $A_x$ of the physical vector potential.   
 The vector-current anomaly therefore describes the force exerted on  the superfluid by  the electric field $E_1$.  
 
In two dimensions, and in the absence of a gauge field,  there is relation between the vector and axial vector Dirac currents in these models:
\bea
J^0_A&=& \frac 1 {v_f} J^1_V,\nonumber\\
J^1_A&=& v_f J^1_V.
\eea
In light of $\sqrt{g}=v_f$, this relation can be written in Lorentz covariant form
\be
J^\mu_A  =-\frac{\epsilon^{\mu\nu}}{\sqrt{g}} J_{\nu, V}
\ee
and  tells us that $f^2=1/4\pi$  and is independent of $v_f$. In higher dimensions $f^2$ will be non-universal.

\section{Four dimensions} 
\label{SEC:four-dimensions}

Having seen that our strategy for obtaining an effective action for Weyl and Dirac fermions coupled to left and right gauge fields gives physically correct  results, we  apply it to the 3+1  dimensional case.

For four dimensions the   Bardeen counterterm is 
\be
S_4(R,L)= \textstyle{\frac 12} \tr\{(LR-RL)(F_R+F_L)+R^3L-L^3R+\frac 12 LRLR\},
\ee
 and 
 the WZW functional becomes 
 \be
 W[R,L,U]=- \frac i{240 \pi^2} \int_M \tr\{(U^{-1}dU)^5\}- \frac{i}{48\pi^2} \int_{\partial M} Z(L,R, U),
 \ee
 where \cite{witten-global,manes,kaymakcalan}
 \bea
 Z(R,L,U)&=& - \tr\{U_L(LdL+dLL+L^3)- U_L^3 L\}-\tr\{R\leftrightarrow L \} \nonumber\\
 &&+\textstyle{\frac 12} \tr\{U_LLU_LL\} - \frac12\{R\leftrightarrow L\}\nonumber\\
 &&-\tr\{U^{-1} LUR^3\} +\tr\{URU^{-1} L^3\}\nonumber\\
 &&- \tr\{U^{-1}LU(RdR+dRR) \}+\tr\{URU^{-1}(LdL+dLL)\}\nonumber\\
 &&- \tr\{URU^{-1} LU_LL\}-\tr\{U^{-1}LURU_RR\}\nonumber\\
 &&+\tr\{LdUU_RRU^{-1}\} +\tr\{RdU^{-1} U_LLU\}\nonumber\\
 &&-\tr\{dLdURU^{-1} \}+\tr\{dRdU^{-1} LU\}\nonumber\\
 &&+\textstyle{\frac 12} \tr\{RU^{-1}LURU^{-1}LU\}.
 \label{EQ:Manes-full}
 \eea
 We are using the  notation $U_L= dUU^{-1}$, and $U_R= U^{-1} dU$ from \cite{witten-global}.

 The rather long and complicated expression  (\ref{EQ:Manes-full}) simplifies greatly in the Abelian  case where  $U=e^{-i\theta}$ because  all terms with more than one $d\theta$ go to zero.  If we then  then set $L=R=A$, we find
 \be
 Z\to 6i \, d\theta\, \tr\left\{AdA+\frac{2}{3}A^3\right\},
 \ee
 making
 \bea
 W[A,\phi]&=& \frac{1}{8\pi^2} \int_{\partial M} d\theta  \,  \tr\left\{AdA+\frac{2}{3}A^3\right\}\nonumber\\
 &=& -\frac{1}{8\pi^2} \int_{\partial M} \theta  \,\tr\{F^2\}.
 \eea
 This is the usual ``$\theta$'' term that appears in topological insulators.

 Now  keep $L$ and $R$ distinct, but make them Abelian.  Then 
 \bea
 \tilde C[R,L,U] &=& \frac{1}{24\pi^2}\int_{M_5} (RF_R^2 -LF_L^2)  \nonumber\\
 &&-\frac{1}{48\pi^2} \int_{M_4} \big\{id\theta(2L\,dL+2R\,dR +RdL +LdR)+ 2LR\,dR-2RL\,dL\big\}.
 \eea
The variation of the Chern-Simons  terms requires knowing 
\be
 \delta \int_{M_5} AF^2 = 3\int_{M_5} \delta A\, F^2  +2\int_{M_4} \delta A \,A F.
 \ee
Varying $R$ gives a surface contribution to the current from 
\be
\delta_R \tilde C[R,L,U] = \frac 1{12\pi^2}\int_{M_4}\delta R\left\{(id\theta -L+R)dR +\frac 12 (id\theta -L+R)dL\right\}.
\ee
Varying $L$ gives 
\be
\delta_L \tilde C[R,L,U] = \frac 1{12\pi^2}\int_{M_4}\delta L\left\{(id\theta -L+R)dL +\frac 12 (id\theta -L+R)dR\right\}.
\label{EQ:second-current}
\ee
Both currents make use of the appropriate  covariant derivative. Some integration by parts is necessary to get these, so there may be extra terms  in  the presence of boundaries or singularities in the $\theta$ field.

There  will also be a  non-topological part of the action such as 
\be
S[\phi] = \int d^4x\left\{ \frac {f^2} 2 (\partial_\mu \theta -L_\mu +R_\mu)(\partial^\mu\theta -L^\mu+R^\mu)\right\},
\ee
where   $f^2$ might be   a superfluid density.
Here we  have again set   
\be
R=iR_\mu dx^\mu
\ee
and similarly for $L$. 

The non-topological action   contributes   currents 
\bea
j^\mu_R &=& -f^2(\partial^\mu\theta-L^\mu+R^\mu),\nonumber\\
j^\mu_L &=&+ f^2(\partial^\mu\theta -L^\mu+R^\mu) 
\eea
that  are to be added to the topological currents found above to make 
\bea
J^\mu_R&=&  -f^2(\partial^\mu\theta-L^\mu+R^\mu)+\frac{1}{24\pi^2} \epsilon^{\mu\nu\sigma\tau}( \partial_\nu\theta-L_\nu+R_\nu)\left(F^R_{\sigma\tau}+\frac 12 F^L_{\sigma\tau}\right),\nonumber\\
J^\mu_L&=&+ f^2(\partial^\mu\theta -L^\mu+R^\mu)+\frac{1}{24\pi^2} \epsilon^{\mu\nu\sigma\tau}( \partial_\nu\theta -L_\nu+R_\nu)\left(F^L_{\sigma\tau}+\frac 12 F^R_{\sigma\tau}\right).
\eea
The equation of motion for the $\theta$ field is
\be
-f^2\partial_\mu(\partial^\mu\theta  -L^\mu+R^\mu)=\frac 1 {96\pi^2} \epsilon^{\mu\nu\sigma\tau}(F^L_{\mu\nu}F^L_{\sigma\tau}+
F^R_{\sigma\tau}F^R_{\mu\nu}+F^R_{\mu\nu}F^L_{\sigma\tau}),
\ee
the RHS coming from the four-dimensional $\theta$  term.

Using the equation of motion gives
\bea
\partial_\mu J^{\mu}_R&=& \phantom- \frac 1{32\pi^2} \epsilon^{\mu\nu\sigma\tau} F^R_{\mu\nu}F^R_{\sigma\tau},\nonumber\\
\partial_\mu J^{\mu}_L&=& -\frac 1{32\pi^2} \epsilon^{\mu\nu\sigma\tau} F^L_{\mu\nu}F^L_{\sigma\tau}.
\eea
We see that, as expected, the coupling to the mass-generating $\theta $ field does not affect the anomaly.

For our chiral superfluid we must set $eA_\mu\equiv R_\mu=-L_\mu$ so that $F^R_{\mu\nu}=-F^L_{\mu\nu}=eF_{\mu\nu}$. We must also  divide by two because of the BdG/Majorana over-counting. The physical particle-number  current for our right-handed Weyl superfluid is therefore  
\be
J^\mu_{\rm Num} 
=- f^2(\partial^\mu\theta +2eA^\mu) +\frac{e}{48\pi^2} \epsilon^{\mu\nu\sigma\tau}( \partial_\mu \theta +2eA_\mu)F^{}_{\sigma\tau}.
\label{EQ:main-current}
\ee
This current has the same anomaly as  a massless right-handed chiral fermion:
\be
\partial_\mu J^\mu_ {\rm Num}= \frac {e^2}{32\pi^2} \epsilon^{\mu\nu\sigma\tau} F_{\mu\nu}F_{\sigma\tau}= \frac{e^2}{4\pi^2}{\bf E}\cdot {\bf B},
\label{EQ:main-EofM}
\ee
so, again as is to be expected,   a  Majorana mass does not affect the anomaly.  Equations (\ref{EQ:main-current}) and (\ref{EQ:main-EofM}) are the principal results of this paper.

 \section{Discussion and application to the chiral magnetic effect}
 \label{SEC:discussion}
 
 By the end of section  \ref {SEC:four-dimensions} we have seen that the  action
 \bea
 S[\theta,A]&=&  \frac 12 \left[\int_{M_4=\partial M_5}  d^4x\left\{ \frac {f^2} 2 (\partial_\mu \theta+2eA_\mu)(\partial^\mu\theta +2eA^\mu)-\frac{\theta}{96\pi^2}\epsilon^{\mu\nu\sigma\tau} F_{\mu\nu}F_{\sigma\tau}\right\} \right] \nonumber\\
 &&\quad -  \frac{1}{96\pi^2}  \int_{M_5} d^5x \epsilon^{\mu\nu\rho \sigma\tau}A_\mu F_{\nu\rho}F_{\sigma\tau}, 
 \label{EQ:final-action}
 \eea
is invariant under the gauge transformation 
 \bea
 \theta &\to& \theta -2\alpha e,\nonumber\\
 A_\mu &\to& A_\mu +\partial_\mu \alpha,
 \eea
 and gives a current  on the $M_4$ space-time boundary of 
 \be
J^\mu_{\rm Num} 
=- f^2(\partial^\mu\theta +2eA^\mu) +\frac{e}{48\pi^2} \epsilon^{\mu\nu\sigma\tau}( \partial_\mu \theta +2eA_\mu)F_{\sigma\tau}.
\label{EQ:4d-number-current}
\ee
This current has the same chiral anomaly 
\be
\partial_\mu J^\mu_{\rm Num}= \frac{e^2}{32\pi^2} \epsilon^{\mu\nu\sigma\tau} F_{\mu\nu}F_{\sigma\tau}
\ee
as the original ungapped  Weyl fermion. One third of the anomaly comes from the topological second term in (\ref{EQ:4d-number-current}) and two-thirds from the non-universal first term and its associated equation of motion 
\be
- f^2\partial_\mu (\partial^\mu\theta +2eA^\mu)= \frac 1{96 \pi^2} \epsilon^{\mu\nu\sigma\tau}F_{\mu\nu}F_{\sigma\tau},
\label{EQ:4d-EofM}
\ee
where the RHS arises from the  $\theta$-term in the $M_4$ part of (\ref{EQ:final-action}).
The space-time current non-conservation  is accounted for by the inflow from the $M_5$ bulk current  described by the last (Chern-Simons) 
 term in (\ref{EQ:final-action}).
In all  these equations $A_\mu$ and $F_{\mu\nu}$ are  the physical  Maxwell fields.  

For a topological superconductor, the  space-time  part of the action  --- the first line in (\ref{EQ:final-action}) ---  describes only one of the two opposite-Chern-number Fermi surfaces.   The second surface will have a similar action, but with the signs of the space-time $\theta$ term and the bulk Chern-Simons term reversed. The resultant cancellation of the $M_5$ terms  ensures that  the complete current that couples to the external electromagnetic field is free of anomalies.

 One may worry that the division between the topological term and the non-universal term  in (\ref{EQ:4d-number-current}) is an artifact of  the ambiguity of the definition of the currents in an anomalous theory --- particularly as a key ingredient in our derivation    
of the effective action involved the Bardeen counter-term which was originally introduced as an ad-hoc modification of the ``consistent'' currents so as to conserve the vector current even in the presence of an axial gauge field.  Bardeen was allowed to add such terms because the AVV and AAA triangle Feynman diagrams  are only {\it conditionally\/}  convergent, and therefore both they and the currents whose response they capture  are  intrinsically ambiguous.  We would argue however that the statement that only one third of the anomaly comes from the  topological term is  unambiguous.  This is because the  coefficient $1/48\pi^2$ of $\partial_\nu\theta$ in the topological part of the current comes from the {\it absolutely  convergent \/}  $\gamma_5 AA$ triangle diagram. It is shown in  appendix \ref{AP:feynman}  that the  $\gamma_5 AA$ diagram evaluates to 1/3 of the absolutely convergent $\gamma_5 VV$ triangle diagram that   gives the corresponding topological part 
\be 
J^\mu_{\rm top}=\frac e{8\pi^2}\epsilon^{\mu\nu\sigma\tau} \partial_\nu \theta F^V_{\sigma\tau}
\ee
of the current for a four-component Dirac particle given  mass by  a neutral Higgs field  \cite{callan-harvey}.   (The extra factor of 1/2 in (\ref{EQ:4d-number-current}) is from the Majorana condition.)

We now have enough information to resolve the paradox described in the introduction.  Recall how it comes about: The total current from the two Fermi surfaces includes a topological term 
\be
J_{\rm top}^\mu=\frac{e}{48\pi^2} \epsilon^{\mu\nu\sigma\tau}( \partial_\nu\theta_+ -\partial_\nu \theta_-)F_{\sigma\tau},
\ee
where $\theta_\pm$ are the order-parameter phases on the $C_1=\pm1$ Fermi surfaces.  Both Fermi surfaces see the same gauge field, so the connection terms in the covariant derivatives have canceled. This current can, however, be non zero when the two order parameters are free to wind independently. In particlular,   
in the presence of a $2\pi$ winding in one of the Fermi surfaces  and an electric field ${\bf E}$ directed parallel to the vortex, we have an inflowing current of 
\be
\dot N = \frac{e|{\bf E}|}{12 \pi} 
\ee
particles per unit length. This inflow is only one-third that found in \cite{qi-zhang-witten}, but is still an embarrassment as the topologically bound mode in the vortex core is uncharged and has no anomaly that can absorb it.   

The resolution of the problem resides in the remark made after equation (\ref{EQ:second-current}) that we might have missed boundary terms arising from integrations by parts in our four-dimensional space-time.  Furthermore there  \underline{will} be boundaries whenever we have a vortex:  We  must exclude from our manifold $M_4$  any   line  about which  $\theta$ winds by $2\pi$ as at such places   $d^2\theta\ne0$.  Finding such boundary  terms by inspection of the algebra is tedious, but we can locate one of them by observing that there can be no physical effect from a singular gauge transformation that is implemented by inserting a half-unit of magnetic  along a line, and simultaneous making $\theta$ wind about the line by $2\pi$ so that  no  covariant derivatives are changed.  Because  our currents  are built from covariant derivatives most of our expressions are unchanged by this process. An exception is the source term $e^2{\bf E}\cdot {\bf B}/{12 \pi^2}$ on the RHS  of the equation of motion for $\theta$. Because a  singular gauge transformation located on the $z$ axis inserts  a flux tube of strength
$(\pi/e) \delta^2(x,y)$,  this source term is modified so that it would appear that charge is being absorbed by the singular line at rate proportional to the component of ${\bf E}$ tangential to the flux line. This cannot be so. We must therefore have missed a compensating source term proportional to $d^2\theta$ that will remain present when $\theta$ winds  but there is no  inserted flux.  This extra term on the RHS of the equation of motion for $\theta$ solves our problem. It provides a source that  under the condition of the paradoxical topological inflow   supplies an equal and opposite outflow in the {\it  non-topological\/}  part of the current.   The net result is that there is no net inflow, and no paradox. 
     

 To  further  illustrate the consequences of  (\ref{EQ:4d-number-current}) we consider the chiral magnetic effect (CME) \cite{CME,CME1,CME2,CME3}  in which a static magnetic field ${\bf B}$ induces a current parallel to the field.  (In our discussion  of the CME   we   consider only  the effect of the {\it external\/}  field  on the superfluid. We are not accounting for any  field generated  by  the currents induced in the condensate. Such  additional geometry-dependent  fields would  lead to a  Meissner effect and tend to screen the fluid from the  external field. Ignoring these screening fields is standard in the usual  derivations  of the CME.)     

Suppose we have a  field ${\bf B}= (0,0,B_3=F_{12})$ that arises from from static  and $x^3$  independent $A_1$, $A_2$ If we allow $A_0$ to depend on $x^3$ or $A_3$ to depend on $x^0=t$,  then anomaly equation becomes 
\be
    \partial_0 J^0_{\rm Num}+\partial_3 J^3_{\rm Num}=\frac 1{4\pi^2}(\partial_0 A_3-\partial_3 A_0) B_3.
\ee 
This  is satisfied by
\bea 
J^0_{\rm Num}&=& \rho_0 + \frac{e^2}{4\pi^2} A_3B_3,\nonumber\\
J^1_{\rm Num}&=&J^2_{Num}=0,\nonumber\\
J^3_{\rm Num}&=&\phantom{ \rho_0}   -\frac{e^2}{4\pi^2} A_0 B_3.
\label{EQ:anomalyCME}
\eea
The last line of   (\ref{EQ:anomalyCME}) leads to the  usual static CME current \cite{CME,CME1,CME2,CME3}
\be 
{\bf J}_{\rm CME}= \frac{e\mu_5 }{2\pi^2} {\bf B}.
\label{EQ:usualCME}
\ee
Here  we have replaced $-eA_0$ by  separate  chemical potentials $\mu_R$, $\mu_L$ for a pair of right- and left-handed Weyl fermions, and then defined  the axial chemical potential $\mu_5$ by setting  $\mu_R =\mu+\mu_5$, $\mu_L =\mu-\mu_5$.

For our superfluid, and when  $B_3 = \partial_1A^2-\partial_2A^1\ne 0$, we  cannot  find a $\theta$ such that
$(\partial^1\theta+2eA^1)= (\partial^2\theta+2eA^2)=0$.   Consequently our formula (\ref{EQ:4d-number-current}) for the current  cannot be coerced to give  (\ref{EQ:usualCME}).  The simplest solution to the  equation of motion for the condensate in the presence of the magetic field   is to take  $\theta$ constant, and this gives us a London-equation-like current 
\bea
J^0_{\rm Num}&=& -2ef^2 A^0,\nonumber\\
 J^1_{\rm Num}&=& -2ef^2 A^1,\nonumber\\ 
 J^2_{\rm Num}&=& -2ef^2 A^2,\nonumber\\
 J^3_{\rm Num} &=&  - \frac {e^2}{12\pi^2} A^0 B_3 =\frac{\mu_Re}{12\pi^2}B_3. 
 \label{EQ:rot-fluid}
 \eea
 Taking $(A^1, A^2)=B_3/2( -y,x)$ and comparing  the $x$, $y$ current components with the number density  $\rho= J^0_{\rm Num}=2f^2 \mu_R$,  we see that this solution corresponds to the  fluid  rotating rigidly with angular velocity
 \be
 {\bm \Omega}= - \left(\frac{e}{2\mu_R}\right) {\bf B},
 \label{EQ:angvelocity}
 \ee
 and   possessing a CME current
 \be
J^3_{\rm Num}  =\frac{\mu_Re}{12\pi^2}B_3, 
 \ee
that  is only 1/3 of the usual equilibrium  value. 
 
 We might   expect some reduction in the strength of the CME because a  degenerate gas of non-interacting Weyl fermions that is rigidly rotating with angular velocity   ${\bm \Omega}$ possesses  an equilibrium  {\it chiral vortical\/} effect (CVE)  current  \cite{CVE, CVE-frequency} of 
 \be
 {\bf J}_{\rm CVE} = \frac{\mu^2}{4\pi^2}{\bm \Omega} .
 \ee
 Given (\ref{EQ:angvelocity}),  this CVE current would  cancel $1/2$ of the usual CME.  We find only  1/3 rather than 1/2 of the noninteracting  CME current remaining, but it is not unreasonable  that the CVE  of a superconductor should differ from  that of the free gas.
 
Is our  rotating solution physically relevant?  Imagine  starting   with  our chiral superfluid at rest and in the absence of any external field.  Now  slowly switch on the  magnetic field. The  circulating electric field from $ {\rm curl\,} {\bf E}=-\dot {\bf B}$ will spin-up the fluid (this is the origin of the London moment of a rotating superconductor)   to   give the  $J^1_{\rm Num}$, $J^2_{Num}$ of (\ref{EQ:rot-fluid}).   Consequently our $\theta=$ const. solution  corresponds to this low-frequency response. 
Finite-frequency computations of the CVE 
(see for example \cite{CVE-frequency} eqs (47,48), or \cite{CME-holography})   show   that the CME current drops from its $\omega=0$ value (\ref{EQ:usualCME}) to exactly one-third of this value as soon as   the frequency $\omega$ becomes non-zero.  It remains at this reduced  value  as long as  $\omega$ is   small  compared to the temperature or the chemical potential.   The physical difference between $\omega=0$ and and  $\omega>0$   in these calculations is that in the former case the fluid has had time    to  relax  to an  equilibrium state in which all  rotational momentum has been shed. Being a  superfluid,  our system will have persistent rotational currents  and  the relaxation time is infinite. Our  result (\ref{EQ:rot-fluid})
applies at both $\omega=0$ and $\omega>0$ and  is  nicely consistent  with the results of   \cite{CVE-frequency,CME-holography}.

 \section{Conclusions}
 
 We have found an action functional, gauge current, and equation of motion for the low energy degrees of freedom  of a Weyl fermion whose superconducting gap (or Majorana mass) is induced by coupling to a charged condensate.  Our expressions for all these quantities differ from those in \cite{qi-zhang-witten}. In particular our expression for the charge current involves  a  topological term that is   smaller by a factor of a one-third than  that in \cite{qi-zhang-witten}. It  also    contains a covariant derivative of the charged order parameter rather than a plain derivative. Despite the coefficient of the topological term being reduced by a factor of  one-third, we find that the chiral anomaly is unchanged. The deficit  is made up of a contribution from a non-universal and non-topological  part of the action  that nonetheless makes a topological contribution through  the influence of the anomaly on the equation of motion obeyed by the Goldstone mode. This additional contribution resolves the threatened  paradox mentioned in section \ref{SEC:Weyl}, where a simple  cosmetic rewrite of the Hamiltonian appears to have reduced the gauge anomaly by a factor of one-third. 
The  contributions  of the non-universal terms to the anomalous effects are independent  of their detailed form so long as they  are conventionally gauge invariant.
 
 In \cite{qi-zhang-witten} the non-universal current   is set equal to the topological current  for reasons that we do not understand, but  
 it is possible that our decomposition of the current into topological and non-topological parts is somehow equivalent to their expression. For example our  distinction between the gauge-covariant derivative and the plain  partial derivative in the topological current  is  insignificant because the connection part cancels when we add the contributions from the two Fermi surfaces.      Full equivalence  seems unlikely, however,  because we have no inflow  into vortex lines,  and because our factor of one-third in the topological current has a real  physical effect of reducing  the CME to  one third of its free equilibrium value.
 
 After this paper was written  we came across a work  \cite{preskill} in which the part of the effective action in (\ref{EQ:final-action}) that arises from the second Fermi surface (including the 1/3 coefficient before  the $\theta$-term) is used to cancel the anomaly of a masless chiral fermion. In \cite{preskill}  the anomaly-cancelling  term is  interpreted as a four-dimensional analogue  of the Green-Schwarz mechanism \cite{green-schwarz} rather than as   a  physical effect  of a gapped  chiral superconductor.

 \label{SEC:conclusions}
 
 \section{Acknowledgements} This  project was supported by the National Science Foundation  under grant    NSF DMR 13-06011.  P.L.S.L. 
 acknowledges support from the  S\~ao Paulo Research Foundation (FAPESP) under grants 2009/18336-0 and 2012/03210-3.  We  would  like to thank  Shinsei Ryu and Jeffrey Teo for discussions and feedback, and also Vatsal Dwivedi for pinpointing   sign errors in an early draft.
 
 \appendix
 \section{Vortex core states}
 \label{AP:vortex}
 Consider a vortex lying along the $z$ axis. 
 The Erick Weinberg index theorem \cite{jackiw-rossi,weinberg-index} guarantees  that when $\Phi$ has winding number $\pm 1 $ there will be a   zero energy solution to the Weyl equation 
\be
H\Psi\equiv \left[\matrix{ {\bm \sigma}\cdot ({\bf p}-e{\bf A}) & \Phi\cr
                                           \Phi^* & -{\bm \sigma}\cdot ({\bf p}+e{\bf A})}\right] \Psi=E\Psi,
 \ee
when  we  restict  ${\bf p}$ and ${\bf A}$  to the  the $x$-$y$ plane.     
If we  consider the rotationally symmetric field    $A_\mu dx^\mu= A_\theta d\theta$ and take  unit positive winding  $\Phi = e^{i\theta}\Delta(r)$ (where $\theta$ is the polar angle, and not  the dynamical Goldstone field)  this  solution is 
\be
\Psi= \left[\matrix{ e^{i\pi/4} \cr 0\cr 0\cr e^{-i\pi/4}}\right]\exp\left \{ - \int_0^r \left(\Delta(\rho) +\frac {A_\theta(\rho)}{\rho}\right) d\rho\right\}.
\ee
If the winding goes the other way $\Phi=e^{-i\theta} \Delta(r)$ it will be 
\be
\Psi= \left[\matrix{ 0 \cr e^{i\pi/4}\cr e^{-i\pi/4}\cr 0}\right]\exp\left \{ - \int_0^r \left(\Delta(\rho) -\frac {A_\theta(\rho)}{\rho}\right) d\rho\right\}.
\ee
Now  we allow motion in the $z$ direction by letting  $\Psi \to e^{ip_3z}\Psi$ and including a gauge field component $A_3$.  This leads to  an additional  term in the Hamiltonian
\be
H(p_3, A_3)= \left[\matrix{ \sigma_3  (p_3-eA_3) & 0\cr
                                           0 & -\sigma_3 (p_3+eA_3)}\right].
 \label{EQ:p_3}                                          
 \ee
 When $A_3=0$ this new operator is  diagonal in the zero-mode basis   with eigenvalue $E(p_3)= + p_3$ in the winding number $+1$ case and $E(p_3)=-p_3$ in the winding number $-1$ case. The zero mode therefore metamorphoses  into a family of  chiral  modes  running up (down) the positive (negative) unit-winding vortex.  
 
 If the coupling to the gauge field were vector-like, the sign before $eA_3$ in the diagonal terms in (\ref{EQ:p_3}) would be the same and  the zero-mode wavefunction would remain  an  eigenstate but with the  energies   shifted by $eA_3$. Then, when   $A_3=-A_z=E_3t$ we would have spectral flow, and hence a 1+1 dimensional  anomaly 
 \be
 \partial_t \rho +\partial_z j_z= \pm \frac{eE_3}{2\pi}.
 \ee
This does not work in the  the axial case (\ref{EQ:p_3})   as the added term is no longer diagonal in the zero-mode basis.  Indeed when we restrict to the zero mode subspace (which is separated by an energy gap from the rest of the two-dimensional spectrum)   the   matrix elements of the perturbation are zero. Consequently,  provided the ${\bf E}$ field is not sufficiently strong as to disrupt the zero mode, the vortex core  modes behave as if they were electrically neutral.   
The vanishing of the matrix elements  actually holds  at second order, and this remains true even  we include   a non-zero chemical potential $\mu=-eA_0$, although the eigenfunctions are more complicated \cite{lopez}.

 \section{Feynman diagrams}  
 \label{AP:feynman}

Here we evaluate  the  triangle  diagrams that determine the coefficients $C_{A,V}$ in the parity violating part of the gradient expansion of vector  and  axial currents  $j^\mu_{V} =C_{V}\epsilon^{\mu\nu\sigma\tau} \partial_\mu \theta  F^{V}_{\mu\nu}$ and   $j^\mu_{A} =C_{A}\epsilon^{\mu\nu\sigma\tau} \partial_\mu \theta F^{ A}_{\mu\nu}$  induced by a  spatially varying Goldstone field and vector and axial-vector gauge fields respectively.  We wish to show that $C_{A}= C_{V}/3$.

We  work in the Euclidean region where the  Feynman integrals are  
\be
I_{\gamma_5VV}^{\mu\nu}(q_1,q_2)= \int \frac {d^4k}{(2\pi)^4} \frac{\tr\{\gamma^5(\slashed{k}+\slashed{q}_1 +m) \gamma^\mu (\slashed{k}+m) \gamma^\nu(\slashed{k}-\slashed{q}_2+m)\}}{((k+q_1)^2+m^2)(k^2+m^2)((k-q_2)^2+m^2)},
\ee
and
\be
I_{\gamma_5AA}^{\mu\nu}(q_1,q_2)= \int \frac {d^4k}{(2\pi)^4} \frac{\tr\{\gamma^5(\slashed{k}+\slashed{q}_1 +m) \gamma^5\gamma^\mu (\slashed{k}+m) \gamma^5\gamma^\nu(\slashed{k}-\slashed {q}_2+m)\}}{((k+q_1)^2+m^2)(k^2+m^2)((k-q_2)^2+m^2)}. 
\ee
Both integrals are convergent.  We only need  to evaluate them for small $q_1$, $q_2$.

We  use  
\be
\tr\{\gamma^5 \gamma^\mu \gamma^\nu \gamma^\sigma \gamma^\tau\}=4 \epsilon^{\mu\nu\sigma\tau}
\ee
to evaluate the traces in the numerators. We find 
 \bea
 &&\tr\{\gamma^5(\slashed{k}+\slashed{q}_1 +m) \gamma^\mu (\slashed{k}+m) \gamma^\nu(\slashed{k}-\slashed {q}_2+m)\}\nonumber\\
 &=&\qquad -4m\,\epsilon^{\mu\nu\alpha\beta} q_1^\alpha q_2^\beta,
 \eea
 and 
 \bea 
 &&\tr\{\gamma^5(\slashed{k}+\slashed{q}_1 +m)\gamma^5 \gamma^\mu (\slashed{k}+m) \gamma^5\gamma^\nu(\slashed{k}-\slashed {q}_2+m)\}\nonumber\\
 &=&- \tr\{\gamma^5(\slashed{k}+\slashed{q}_1 +m) \gamma^\mu (-\slashed{k}+m) \gamma^\nu(\slashed{k}-\slashed {q}_2+m)\}\nonumber\\
 &=&\qquad+ 4m\,\epsilon^{\mu\nu\alpha\beta} (q_1^\alpha q_2^\beta+2 k^\alpha q_2^\beta - 2q^\alpha_1 k^\beta).
 \eea

We need the standard  integrals  
\be
\int \frac {d^4k}{(2\pi) ^4} \frac{1}{(k^2+m^2)^3}= \frac{1}{32 \pi^{2}} \frac{1}{m^2},
\ee
 \be
\int \frac {d^4k}{(2\pi) ^4} \frac{k^\mu k^\nu}{(k^2+m^2)^4}= \frac{1}{32 \pi^2} g^{\mu\nu}\frac{1}{m^2}\frac{1}{6},
\ee
from which we obtain 
\be
\int \frac {d^4k}{(2\pi) ^4} \frac{1}{((k+q_1)^2+m^2)(k^2+m^2)((k-q_2)^2+m^2) }=  \frac{1}{32 \pi^{2}} \frac{1}{m^2}+O\left(\frac{|q|^2}{m^2}\right),
\ee
and 
\be
\int \frac {d^4k}{(2\pi) ^4} \frac{k^\alpha }{((k+q_1)^2+m^2)(k^2+m^2)((k-q_2)^2+m^2) }= -\frac 13 (q_1^\alpha-q_2^\alpha ) \frac{1}{32 \pi^{2}} \frac{1}{m^2}
+O\left(\frac{|q|^2}{m^2}\right).
\ee
 The last integral leads to the substitution  
 \be
 k^\alpha \to -\frac 13 (q_1^\alpha -q_2^\alpha)
 \ee
 in the second trace, and  gives 
 \be
 I_{\gamma_5AA}^{\mu\nu}(q_1,q_2)=\frac 13 I_{\gamma_5VV}^{\mu\nu}(q_1,q_2)
 \ee
 for small $q$.

\end{document}